# *Low-Temperature Synthesis of Weakly Confined Carbyne inside Single-Walled Carbon Nanotubes*

Bo-Wen Zhang, Xi-Yang Qiu, Yicheng Ma, Qingmei Hu, Aina Fitó-Parera, Ikuma Kohata, Ya Feng, Yongjia Zheng, Chiyu Zhang, Yutaka Matsuo, YuHuang Wang, Shohei Chiashi, Keigo Otsuka, Rong Xiang, Dmitry I. Levshov, Sofie Cambré, Wim Wenseleers, Slava V. Rotkin, and Shigeo Maruyama [*]

**ABSTRACT**: Carbyne, a one-dimensional (1D) carbon allotrope with alternating triple and single bonds, has the highest known mechanical strength but is unstable to bending, limiting synthesis to short linear chains. Encapsulation within carbon nanotubes (CNTs) stabilizes carbyne, forming confined carbyne (CC), thus enabling further research concerning attractive 1D physics and materials properties of carbyne. While CC has been synthesized in multi-walled CNTs (MWCNTs) using the arc-discharge method and in double-walled CNTs (DWCNTs) via high-temperature high-vacuum (HTHV) method, synthesis in single-walled CNTs (SWCNTs) has been challenging due to their fragility under such conditions. In this work, we report a low-temperature method to synthesize CC inside SWCNTs (CC@SWCNT). By annealing SWCNTs containing ammonium deoxycholate (ADC) at 400°C, ADC is converted into CC without damaging the SWCNTs. Raman spectroscopy revealed a strong CC phonon (CC-mode) peak at 1860–1870 cm$^{-1}$, much stronger than the SWCNT G-band peak, confirming a high fraction of CC in the resulting material. The Raman mapping result showed the uniformity of the CC-mode signal across the entire film sample, proving the high efficiency of this method in synthesizing CC in every SWCNT of appropriate size. Notably, the CC-mode peaks of CC@SWCNT (above 1860 cm$^{-1}$) are higher than those reported in previous CC@CNT samples (mostly <1856 cm$^{-1}$). This is attributed to larger SWCNT diameters (>0.95 nm) used in this study, compared to the typical 0.6–0.8 nm range. Larger diameters result in reduced confinement, allowing carbyne to closely resemble free-standing carbyne while remaining stabilized. This low-temperature synthesis of long-chain, nearly free-standing carbyne within large-diameter SWCNTs offers new opportunities for exploring 1D physics and the unique properties of carbyne for potential applications.

**INTRODUCTION**

Carbyne, a one-dimensional (1D) carbon allotrope, consists of a quasi-infinite chain of sp-hybridized carbon atoms. The carbon orbitals can form linear-chain bonds with neighboring atoms in either single- or double- or triple-bond configuration, responsible for the two isomeric structures of carbyne. While the polyyne chain consists of alternated single- and triple bonds, cumulene shows identical double bonds. The existence and degree of bond alternation determine the electronic structure of the carbyne. Polyyne acts as a Peierls semiconductor while cumulene behaves as a metal [1,2]. The so-called bond length alternation (BLA) is used to quantify the difference between the allotropes and describe the difference between the neighboring bond lengths. For cumulene, the BLA is equal to 0, while for polyynes the BLA is larger than zero [3]. Since 1D metal is unstable under the Peierls' distortion [4,5] the polyyne material is considered to be the ground state. This instability is reflected in the strong structure-property dependence, for example, the band gap energy which grows with increasing BLA, and thus being tunable by strain [6,7,8] as well as temperature [9]. Rich many-body physics was theoretically proposed for this 1D material with strong interactions. The sp-hybridized carbon chain structure endows carbyne with excellent mechanical properties: based on theoretical calculations, a tensile strength exceeds 250 GPa [10,11], specific tensile strength is around $7.5\times10^7$ N·m/kg [6], and Young's modulus is around 32.7 TPa [6], stronger than any known material. Additionally, the high frequency of lattice vibrations and the long mean free path of phonons result in exceptionally high thermal conductivity, expected to reach approximately 80 kW/(m·K) at 200–300 K [12]. These excellent physical properties make carbyne a promising candidate for applications in nanoelectronics, spintronic devices, and high-strength composite materials [13,14,15].

However, due to the high reactivity of sp-carbon, in ambient conditions, closely spaced carbyne chains easily bend, couple, and convert into sp2-hybridized carbon structures through cross-linking reactions, making it difficult to synthesize free-standing carbyne. To stabilize the structure, it was necessary to introduce functional groups, such as tert-butyl [16], triethylsilyl [17], complex platinum(II) compounds [18], and other functional groups with high steric hindrance, at both ends to increase the spacing between carbyne chains. In 2010, W. A. Chalifoux *et al.* successfully synthesized carbon chains with 44 contiguous carbon atoms by using the sterically hindered tris(3,5-di-t-butylphenyl)methyl as end groups through organic chemical reactions [19]. More recently, utilizing surface synthesis strategies, W. Gao *et al.* have extended the length of free-standing carbon chains to approximately 120 carbon atoms [20]. Nevertheless, achieving bulk synthesis of carbon chains by introducing large end groups through organic chemical reactions remains challenging. As an alternative, encapsulating carbon chain segments in carbon nanotubes (CNTs) to form confined carbyne (CC) for stabilization has been developed in recent years [21].

Compared to organic chemical reactions, not only is this approach simpler and suitable for large-scale synthesis, but also makes much longer stable chains. In 2003, X. L. Zhao *et al.* synthesized CC consisting of more than 100 carbon atoms, inside multi-walled carbon nanotubes (MWCNTs) using the arc-discharge method [22]. Later, in 2016, L. Shi *et al*. reported a significant advancement by synthesizing ultra-long CC chains, containing over 6000 carbon atoms, within double-walled carbon nanotubes (DWCNTs) under high-temperature high-vacuum (HTHV) conditions (900°C–1460°C) [23], referred to as a linear carbon chain (LCC) or long linear carbon chain (LLCC) in the previous studies. Properties of a linear carbon chain become length-independent when it reaches a certain length and resemble those of infinite carbyne, therefore given the name confined carbyne (CC). However, the HTHV heating treatment method is limited to DWCNTs, as the synthesis of CC@DWCNT requires temperatures exceeding 900°C. Such conditions pose a challenge for single-walled carbon nanotubes (SWCNTs) [24,25], which struggle to maintain stability at these high temperatures [23,26], making it difficult to apply this method to synthesize CC@SWCNT. CC@SWCNT possesses the simplest structure compared with CC@DWCNT and CC@MWCNT. Moreover, CC@SWCNT is predicted to potentially exhibit superconductivity and ferromagnetism [27], as well as possess a significantly larger elastic modulus compared to isolated CC [28]. Yet, to date, no efficient method for synthesizing CC@SWCNT has been reported.

To overcome these challenges, researchers have explored alternative approaches, such as filling open-ended SWCNTs with polyynes ($C_{2n}H_2$) [29,30] or organic solvent molecules [31] as precursors for CC and performing heat treatments at lower temperatures (700°C and 800°C). Alternatively, CC@SWCNT can also be obtained by extracting the inner tube of CC@DWCNT through tip-ultrasonication and density gradient ultracentrifugation, though with yields that remain a challenge to optimize [32]. Despite these efforts, the synthesis efficiency of CC@SWCNT has shown room for improvement. Herein, we report a method utilizing surfactant molecules as the CC precursor. By choosing the surfactant's molecular structure to lower its decomposition temperature, we synthesized CC@SWCNT at 400°C with high efficiency (Fig. 1). To our knowledge, this is the lowest temperature reported for CC@CNT synthesis to date.

Moreover, the CC@SWCNT samples we obtained exhibited a CC-mode Raman shift in the 1860-1870 cm$^{-1}$ range. Based on the previously established relationship between the Raman shift of CC mode and CNT host diameter [33] and our molecular simulation result, the diameter of our host SWCNTs exceeds 0.95 nm. This is larger than the optimal diameter of approximately 0.7 nm (reported in previous studies [23]. Unlike the small-diameter CNT host (6,5) or (6,4), where strong interactions exist between the internal CC and the walls of CNT host [23], our large-diameter SWCNT host results in a much weaker perturbation of the CC by the SWCNT host. This can be confirmed by the nanotube Raman spectrum. The G-band shape observed before and after the formation of the inner CC shows no "deformation" due to their interactions, and the normalized CC-mode intensity ($I_{CC}/I_{G+}$, calculated using peak heights) remains unchanged at different temperatures. The weak interaction potential induced by the large-diameter SWCNT host places the CC in a less confined state, thus less impacting the natural BLA of carbyne (less strain) and thereby resulting in a higher Raman frequency. In this work, while the CC is fully stabilized by the SWCNT host, it is less confined and approaches more the state of free-standing 1D carbyne. Therefore, we have named it "weakly confined carbyne" (wCC). This method is efficient and convenient, filling the gap in synthesizing CC inside large-diameter SWCNT hosts at low temperatures using surfactant molecules as precursors and benefiting subsequent studies on the properties of carbyne in nearly ideal environments.

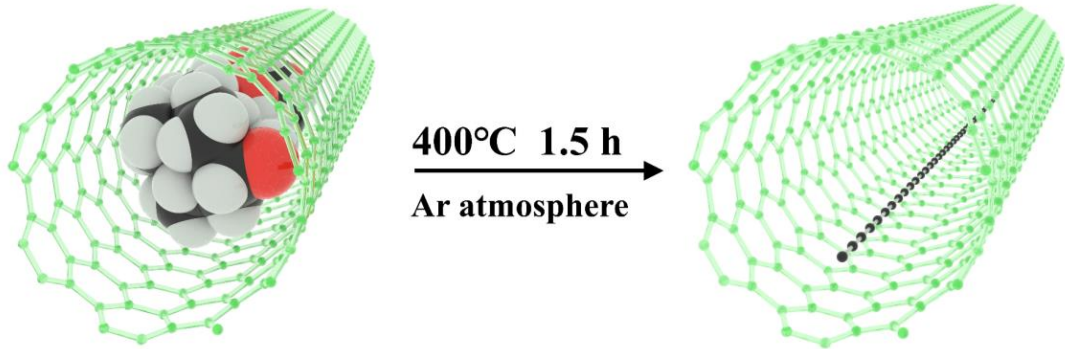

Figure 1. Schematic of CC@SWCNT synthesis: ADC molecule (atomic size defined by the van der Waals radius) insertion into an (8,7) SWCNT and carbyne formation at 400°C.

## RESULT AND DISCUSSION

The synthesis process of CC@SWCNT can be summarized as follows: HiPco SWCNT powder (NanoIntegris, Batch# R2-172) was dispersed by the tip-sonication method, followed by liquid chromatography method to separate semi-conducting and metallic SWCNT dispersions [34]. The dispersion was then subjected to ultrafiltration to replace the surfactant with ammonium deoxycholate (ADC). Using the SWCNT-ADC dispersion, filtration was performed with an anodic aluminum oxide (AAO) membrane according to the method in the previous work [35] to obtain a SWCNT film. Finally, the SWCNT film is annealed at 400°C for 1.5 hours under an argon atmosphere (300 sccm) to synthesize CC@SWCNT. The presence of CC@SWCNT was confirmed through Raman spectroscopy. As shown in Figure 2, the Raman spectra obtained using a 532 nm excitation wavelength revealed the typical radial breathing mode (RBM), D band, G band, and 2D band characteristics of SWCNT before annealing. After annealing, a new sharp peak emerges at around 1863 cm$^{-1}$, with an intensity several times greater than that of the G band. This peak, referred to as the CC-mode peak, is attributed to the stretching vibrations of the carbon-carbon triple bonds, with the peak position depending on the exact value of the BLA [36,37,38,39] (C≡C). Additionally, another new peak appeared at around 3700 cm$^{-1}$, termed the overtone of the CC mode, *i.e.* the 2CC peak [29,40]. The appearance of these two peaks indicates the existence of the CC within the SWCNTs. Note that the CC-mode has previously been shown to have a very high Raman cross-section, up to 3 orders of magnitude higher than the SWCNT RBMs [41], strongly depending on the resonance conditions of both. Though such a strong peak intensity is not surprising, we emphasize that, compared to other bulk measurements of CC in DWCNTs [42], the relative intensity concerning the RBMs and G-band is much higher, therefore demonstrating a large fraction of carbyne material formed and a very efficient growth.

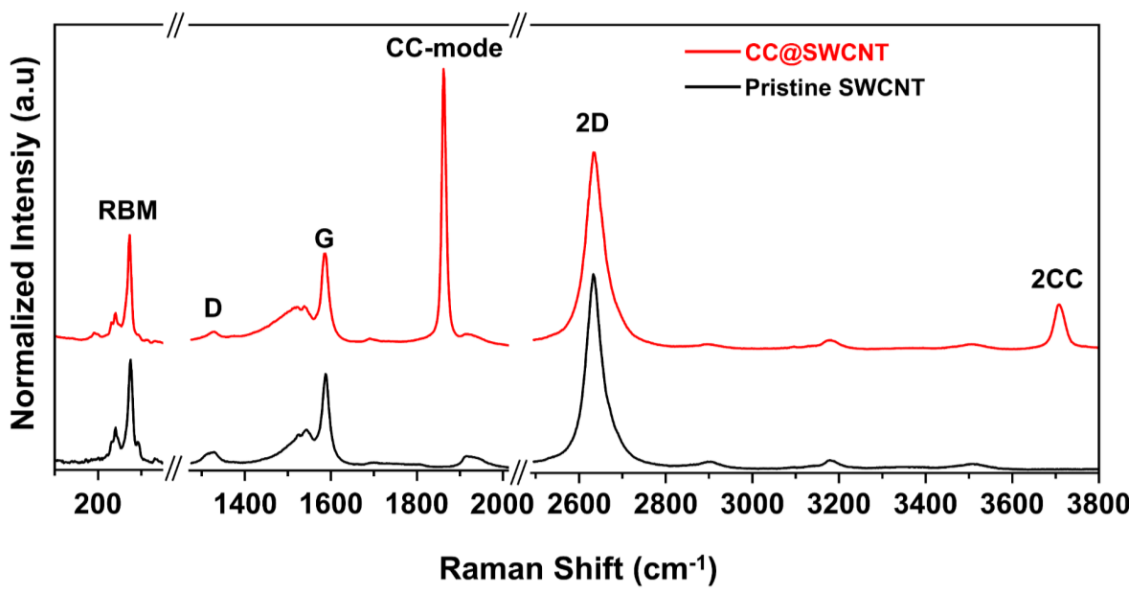

**Figure 2.** Raman spectra of CC@SWCNT synthesized by annealing a SWCNT film containing ADC surfactant under 400°C for 1.5 hours (red curve) and pristine SWCNT (black curve), obtained from a single-point measurement in a Raman imaging microscope setup (but exciting multiple SWCNTs at the same time as shown by the RBMs).

In addition to the 532 nm laser excitation, the resonance Raman conditions were examined at other wavelengths: 488 nm, 633 nm, and 785 nm for CC@SWCNT (Fig. 3). Our samples also displayed the CC-mode peak at the higher laser energy of 488 nm, while the CC-mode peak disappeared for the lower laser energies of 633 nm and 785 nm. We conclude that the resonance wavelength of our CC@SWCNT samples is closer to 532 nm (2.33 eV) by comparing the peak intensities at 488 nm and 532 nm. As earlier studies suggest, the higher the BLA, the higher the energy of the Peierls' band gap, thus the differences in chain length and local strain (as well as temperature) may lead to changes of the band gap energy and the Raman scattering resonance conditions [23,43,44]. Earlier studies demonstrated a strong linear correlation between the CC mode Raman frequency and the resonant laser energy [44]. We found our data (a CC mode peak of 1863 cm$^{-1}$ and near-resonant laser energy of 2.33 eV (532 nm)) consistent with the previously established linear correlation [44] (Fig. S1). The slight deviations arise primarily because our testing laser wavelength at 532 nm was not at the precise resonant wavelength. According to the linear fitting equation, the resonant laser energy for our CC@SWCNT should be slightly less than 2.33 eV, corresponding to a wavelength of 542 nm to be verified in the future with wavelength-dependent Raman spectroscopy, beyond the scope of this work.

Additional evidence for the weak interaction between the confirmed carbyne and the host SWCNT is obtained by comparing the Raman spectra of pristine SWCNTs with those of CC@SWCNTs. In contrast to CC@DWCNT materials synthesized under HTHV — where a strong interaction between CC and CNT host [23,44,45] leads to noticeable changes in the Raman spectrum, such as modifications in shapes and positions of the RBM, G, and 2D peaks, as well as the appearance of new peaks in the G and 2D bands [46] — our CC@SWCNT samples show no significant changes in the Raman spectrum. Under 532 nm excitation, we note only slight frequency shifts in the RBM, $G^-$, and $G^+$ peaks, while under the 633 nm excitation, only a shift in the RBM peak is detected. To elucidate this phenomenon, we compared the Raman mapping results of pristine SWCNTs, annealed SWCNTs, and CC@SWCNTs (Fig. S2a-d). The statistical analysis indicates that both annealed SWCNTs and CC@SWCNTs show similar shifts compared to pristine SWCNTs. This suggests that the peak shift observed in our CC@SWCNT samples is primarily attributed to the annealing process itself, most likely related to bundling/debundling of the SWCNTs before/after annealing (Fig. S3), rather than the interaction between inner CC and host SWCNTs. In neglecting these shifts, the absence of changes in nanotube Raman spectra upon the synthesis of CC inside SWCNT suggests that the interaction between CC and d SWCNT is weak. Since in our samples the confined carbyne is in a relatively non-interacting state, we refer to this as "weakly confined carbyne" (wCC).

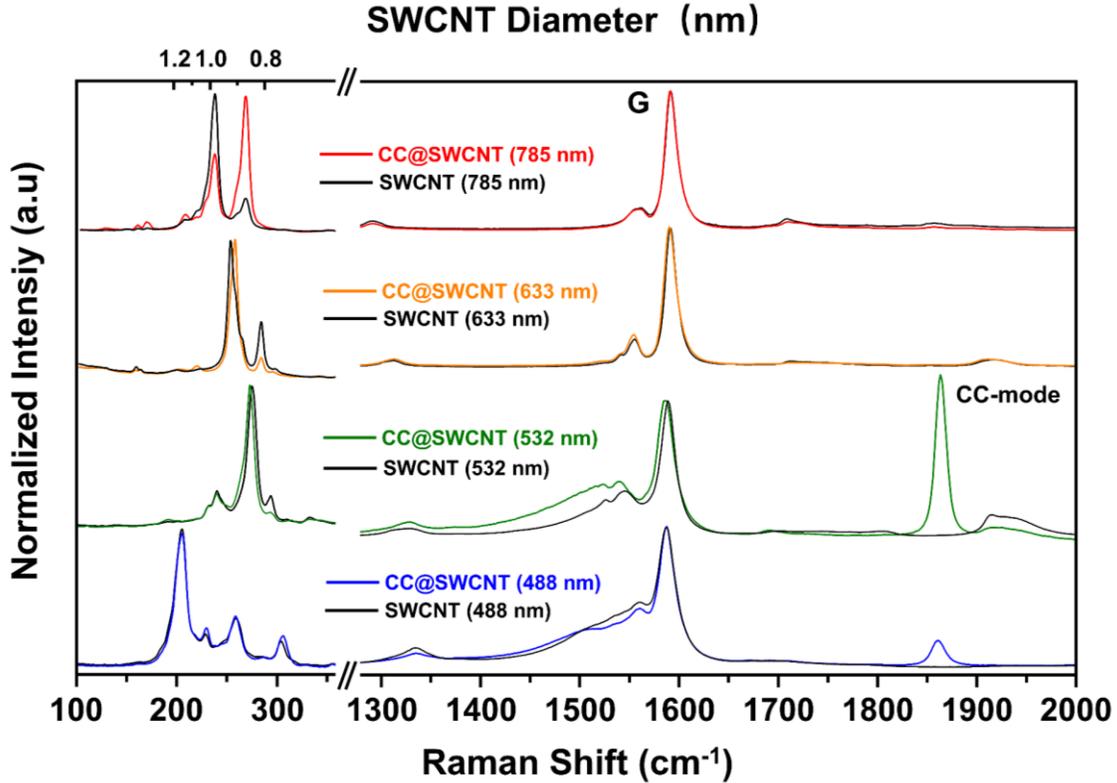

**Figure 3**. Raman spectra of CC@SWCNT and pristine SWCNT (black curve) characterized by 488, 532, 633, and 785 nm excitation wavelengths. The G band and CC-mode regions (right-hand side) are normalized to the intensity of the $G^+$ peak, while the RBM region (left-hand side) is normalized to the highest intensity within the RBM peaks.

To further quantify the amount of interaction between weakly confined carbyne and SWCNTs in our samples, temperature-dependent Raman spectroscopy was introduced (Fig. 4a). Next, we analyze the CC-mode intensity and FWHM under different temperatures. In previous studies, the van der Waals (vdW) coupling between the inner tubes and the CC was shown to depend on the temperature, leading to variations in the CC-mode intensity and FWHM [23]. We measured the Raman spectra of the CC@SWCNT sample at 12 different temperatures, ranging from 78K to 358K, and found that the relative CC-mode intensity and line width did not exhibit significant changes with the temperature variation. The pronounced trend of decreasing peak intensity (intensity ratio $I_{CC}/I_{G+}$ decreased by the order of magnitude) and increasing FWHM (more than twice) with temperature increasing from 50K to R.T. observed in CC@DWCNT [23] (black curves) is in significant contrast to nearly constant peak intensity and slightly increasing FWHM in our samples (Fig. 4b). This demonstrates that temperature variations have no significant impact on the van der Waals coupling between the SWCNT and CC. The reason lies in the fact that the interaction between CC and SWCNT in our sample is already weak, making it nearly unaffected by temperature, which explains the absence of changes in the CC-mode peak intensity and line width. We argue that this weak interaction is due to the larger diameter of the SWCNTs, which will be discussed in detail later. Meanwhile, the observed peak position of the CC mode demonstrated the frequency blue shift as the temperature decreased (Fig. 4c), which is expected due to the decrease of the BLA with the temperature, consistent with the result of CC@CNT (Fig. 4c) in previous papers [23,47].

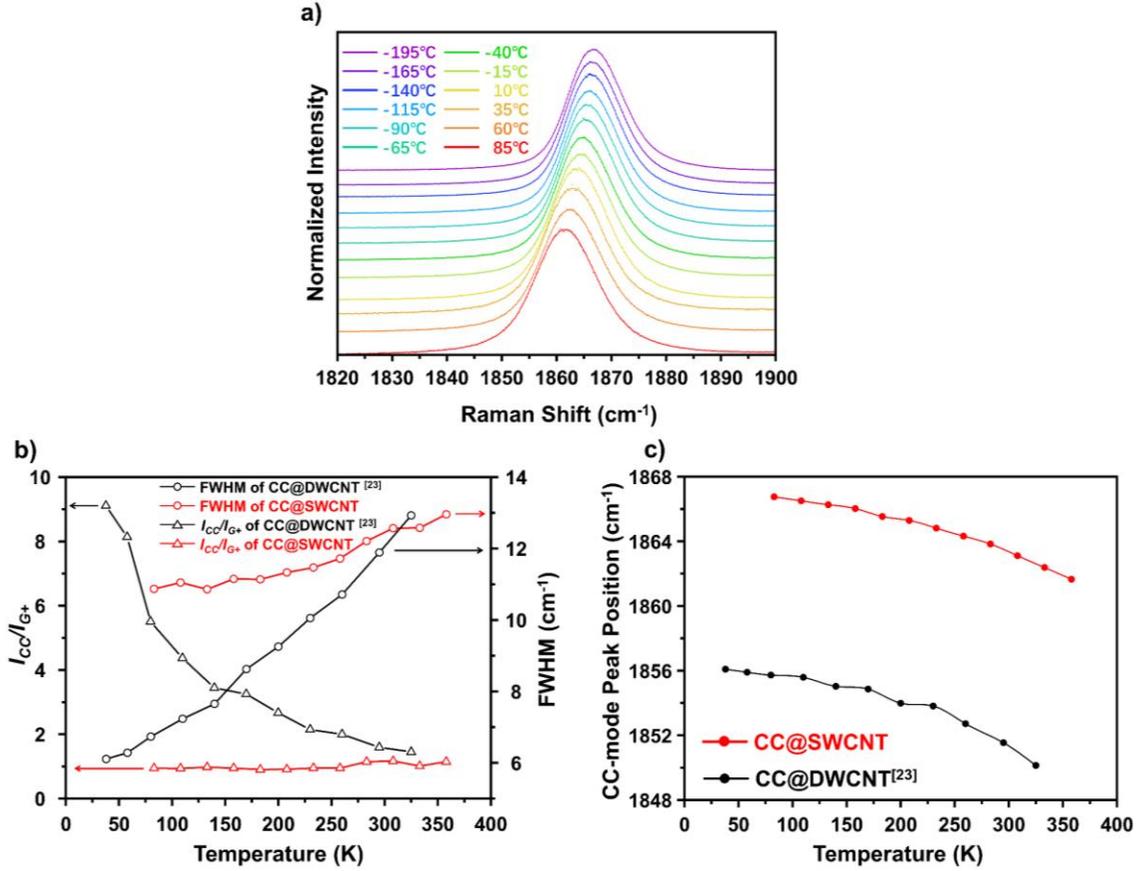

**Figure 4**. (a) Raman spectra of CC@SWCNT measured at different temperatures under 532 nm excitation wavelength. The spectra are normalized to the G$^+$ band intensity. (b) Plot of FWHM and normalized CC-mode intensity ($I_{CC}/I_{G+}$) as a function of temperature. Red triangles and circles represent the $I_{CC}/I_{G+}$ and FWHM of CC@SWCNT, while black triangles and circles represent the $I_{CC}/I_{G+}$ and FWHM of CC@DWCNT from the previous paper [23]. (c) The CC-mode peak position as a function of temperature. Red and black circles represent the CC@SWCNT and CC@DWCNT from the previous paper [23].

Next, we will discuss the efficiency of CC synthesis. To examine the formation of CC at different locations over the SWCNT film, we performed Raman mapping to scan the entire film sample. One needs to consider the natural non-uniformity of the pristine SWCNT film. Thus, we used the normalized CC-mode intensity, $I_{CC}/I_{G+}$, as in previous works, to evaluate the carbyne content via the intensity of the CC-mode signal at various points. As shown on the Raman mapping in Figure 5a, the CC-mode signal is present across the entire sample for successful synthesis. A histogram of the $I_{CC}/I_{G+}$ data shows a narrow distribution of $I_{CC}/I_{G+}$ values in the range of 1–5 (Fig. 5b). The distribution exhibits an approximately normal shape, symmetric around the peak at 3. This indicates that at most of the sample's locations, CC-mode intensity is at least of the order of, and in many locations, several times greater than the G-band intensity, demonstrating the high synthesis efficiency. An absolute measure of efficiency is hard to provide based on Raman measurements. One should take into account carefully the previously determined Raman cross-section for the G- and CC bands, as well as the resonance conditions that vary for different nanotubes. Moreover, not all SWCNTs in the present sample have sufficiently large diameters for CC growth. Still, assuming some of the SWCNTs are not filled and some are not in the resonance to give a strong enough Raman signal, we note that the typical ratio $I_{CC}/I_{G+}$ was of the order of 1 and less in the previous studies [42,44].

It should be noted that this synthesis of CC@SWCNT exhibits some variability. We repeated the growth under similar growth conditions (identical parameters) and characterized seven different CC@SWCNT samples, named SAMPLE A to G, with SAMPLE G being the best sample presented above. The Raman mapping demonstrated a gradient-like distribution of CC in samples A to C around the edges of the studied area 100×100 um$^2$, while in samples D to F, it gradually spread with a maximum toward the center. Sample G showed the most uniform CC-mode signal, covering the entire area studied (Fig. S4a-Fig. S4g). From the $I_{CC}/I_{G+}$ distribution histograms, a more uniform growth process corresponds to the shift of the distribution maximum to the higher $I_{CC}/I_{G+}$ value, accompanied by a broader distribution (Fig. S4h). The average Raman spectra are shown in Fig. S5. After normalizing the G$^+$ band intensity, an apparent increase in the CC-mode intensity can be observed from less uniform and less efficient growth to the most uniform and most efficient sample (SAMPLE A to SAMPLE G). The reason for the variation in the CC-mode intensity is not yet clear at this moment. Considering that the filtration process of the SWCNT film involves manual deposition, each SWCNT film sample is inevitably different. In some locations, the ADC might be washed away through the filtration process. These differences in the synthesis conditions may lead to variations in the CC-mode peak intensity among different samples, and might be due to the collection of non-uniform Raman signal itself, calling for further investigation to clarify the specific details.

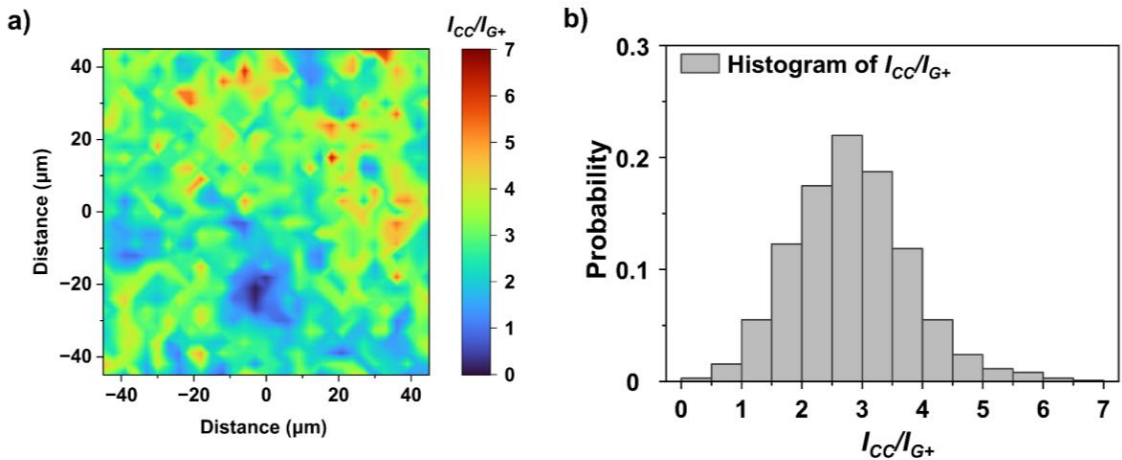

**Figure 5**. (a) Raman mapping of normalized CC-mode intensity ($I_{CC}/I_{G+}$) of the best CC@SWCNT sample (SAMPLE G) characterized by 532 nm laser wavelength (mapping area 45 μm × 45μm, step 3μm). (b) Histogram of $I_{CC}/I_{G+}$ of the best CC@SWCNT sample (SAMPLE G).

Understanding the synthesis mechanism of CC@SWCNT by annealing SWCNT containing ADC surfactant molecules is critical for further improving and stabilizing its efficiency. We believe this process is similar to the previously reported method, where SWCNTs were filled with polyynes as CC precursor (as opposed to surfactant molecules in our study) and annealed to synthesize CC@SWCNT [29]. To verify this assumption, we conducted synthesis experiments using different surfactant molecules, such as sodium cholate (SC), sodium deoxycholate (DOC), sodium dodecyl sulfate (SDS), and ammonium deoxycholate (ADC). The results showed that all annealed samples exhibited the CC-mode. It should be noted that similar to the variation observed from SAMPLE A to SAMPLE G in CC@SWCNT synthesized using ADC, the CC-fraction also varied for each surfactant (SC, DOC, and SDS). For each precursor, we conducted multiple experiments (synthesis followed by collecting the average Raman spectrum for the sample with the highest $I_{CC}/I_{G+}$) and selected the results for comparison in Fig. S6a-c. Among the sodium-containing surfactants (DOC, SC, and SDS), the sample obtained from the DOC dispersion exhibited the highest average CC-mode peak intensity, the broadest distribution of CC-mode peaks, and the most uniformity in the Raman mapping (Fig. S6e-f). However, even for DOC, the results still lag far behind the best synthesis achieved with ADC (Fig. S6d). We speculate that by replacing the sodium cation of DOC with the volatile ammonium cation, the ADC surfactant enhanced growth efficiency by lowering the thermal decomposition temperature, making it more prone to decompose into carbonaceous residues under thermal treatment [48]. Since the transition from SC, DOC, SDS, and ADC to CC was observed, we consider formation might also occur with encapsulating other organic molecules. We will investigate this transformation in our future research.

Finally, to evidence the size of the surfactant molecules versus the SWCNT diameter plays a role in the synthesis of the CC@SWCNT, we studied SWCNT templates with different SWCNT diameters. Three SWCNT templates were used: HiPco SWCNT and SWCNTs synthesized via the alcohol catalytic chemical vapor deposition (ACCVD) at 600°C and 850°C [49,50]. The diameter distribution of the ACCVD SWCNT can be controlled by adjusting the synthesis temperature [51]. Specifically, the lower temperature (600°C) results in smaller SWCNT diameters, while the higher temperature (850°C) leads to larger diameters. The experimental results showed that the CC-mode signal was detected in both HiPco and 850°C ACCVD SWCNT samples (Fig. 5a, Fig. S7a-b) but not in the 600°C ACCVD SWCNT sample with the smaller average diameter of nanotubes (Fig. S7b). A comparison of the chirality distributions, acquired by photoluminescence excitation (PLE) mapping and fits of those PLE maps (Fig. S8) [52], revealed that the HiPco and 850°C ACCVD SWCNT samples contained a large fraction of components with diameters exceeding 0.95 nm, while the 600°C ACCVD SWCNT sample was primarily composed of components with diameters ranging from 0.75 to 0.95 nm. Based on the PLE and Raman results, the threshold SWCNT diameter for the successful formation of CC@SWCNT is approximately 0.95 nm. This diameter dependency may be related to the size of the surfactant molecules to fill SWCNT and, later, convert into CC. Moreover, the peak position of the CC-mode is also consistent with the fact that the larger-diameter SWCNTs (over 0.95 nm) are filled in our samples. Previous studies have utilized tip-enhanced Raman scattering (TERS) spectroscopy to obtain RBM and CC-mode data for individual CNT, thereby concluding a linear relationship between the Raman peak position of CC-mode ($\omega_{cc}$) and the diameter of the CC host (d), described by the equation $\omega_{cc}$ (cm$^{-1}$) = 1487 + 462d (nm) (0.626 nm<d<0.747nm) [33]. Experimentally, the CC-mode resonance peaks were found at 1772, 1792, 1798, 1831, and 1835 cm$^{-1}$ within the linear region [33], also outside the linear region at 1842, 1850, and 1856 cm$^{-1}$ [44], which corresponds with the CNT host diameter of 0.618 to 0.799 nm according to calculations. The calculation results are consistent with the actual observations from the Transmission Electron Microscopy (TEM), where most CCs have been found inside

CNTs with diameters ranging from 0.62 nm to 0.85 nm [23,31,33]. In previous studies, a CC-mode exceeding 1860 cm$^{-1}$ has been rarely observed due to the limited diameter range of the CC host. However, in our CC@SWCNT samples, the average peak position of CC-mode is distributed within the range of 1863-1867 cm$^{-1}$, as shown by the red curve in Figure 6. This indicates that the SWCNTs accommodating the CC must have larger diameters than in the previous experiments with CNTs, significantly exceeding the theoretically most stable diameter of 0.7 nm calculated by DFT [23]. Since the SWCNT film is a mixture of various chiralities, it would be challenging to precisely determine the diameter of the SWCNTs encapsulating a particular CC linear chain. However, based on the PLE mapping of ACCVD SWCNT samples, we can conservatively conclude that the SWCNT diameters necessary for synthesizing CC@SWCNT from ADC should exceed 0.95 nm (red diagonal hatched region in Fig. 6). Furthermore, when correlating the CC-mode (1863–1867 cm$^{-1}$) of CC@SWCNT samples with CNT diameter (over 0.95 nm), we find that the linear relationship breaks down in the larger diameter range, displaying a dashed curve as shown in Figure 7, which is likely due to the saturation of decreasing CC-host interaction at the limit of quasi-free-standing CC. This weaker interaction between the CC and the CNT wall leads to a weakly confined carbyne state, which explains the lack of deformation in the G-band shape (Fig. 3) and the consistency in the CC-mode peak intensity and FWHM across different temperatures (Fig. 4b).

To better understand the process of CC synthesis inside large-diameter SWCNT, we performed molecular simulations using Avogadro (Version 1.2.0) [53] under the MMFF94 force field [54] with an open boundary condition. 4nm-long CC (28 carbon atoms) terminated with hydrogen atoms, and ADC molecules, where the ammonium was substituted with hydron, were positioned inside SWCNTs of varying diameters. The resulting total energies were compared to those when the molecules and SWCNTs were separated. Based on the energy differences between the "inside" and "separated" configurations, ADC molecules preferentially insert into SWCNTs with diameters exceeding 0.95 nm, as evidenced by the energy approaching zero at a (7,7) chirality diameter of 0.962 nm and turning negative with further increases in diameter (Fig. 7), which is consistent with the PLE and chirality mapping results (Fig. S8). Meanwhile, simulation results indicate that for larger SWCNTs, CCs tend to stick to the inner walls of the nanotubes. Starting from SWCNTs with diameters around 0.8 nm, the CCs shift away from the center and approach the edges, resulting in weaker interactions between the CCs and the SWCNTs (Fig. S9). Thus, a non-linear relationship between SWCNT diameter and CC Raman shift is expected for larger tubes (Fig. 7) instead of a linear correlation. This suggests that the small-diameter nanotubes do not allow the insertion of ADC molecules.

In addition to the SWCNT diameter, we found another critical parameter for the synthesis of CC@SWCNT being the annealing temperature. For the ADC-SWCNT samples, we performed a rough temperature optimization experiment (300°C, 350°C, 400°C, 500°C). It was observed that when the annealing temperature was lowered to 350°C, which is below the thermal decomposition temperature of ADC [48], the CC-mode peak disappeared (Fig. S10). This suggests that the formation of CC is related to the decomposition of surfactant molecules and further supports the hypothesis that surfactant molecules act as the precursor for CC. Meanwhile, we found that when the annealing temperature increases to 500°C, the intensity of the CC mode decreases compared to that at 400°C (Fig. S10). Although we are not yet certain why raising the temperature reduces synthesis efficiency, it may be related to the decomposition and reassembly processes of ADC in forming CC, which will be studied and illustrated in future research.

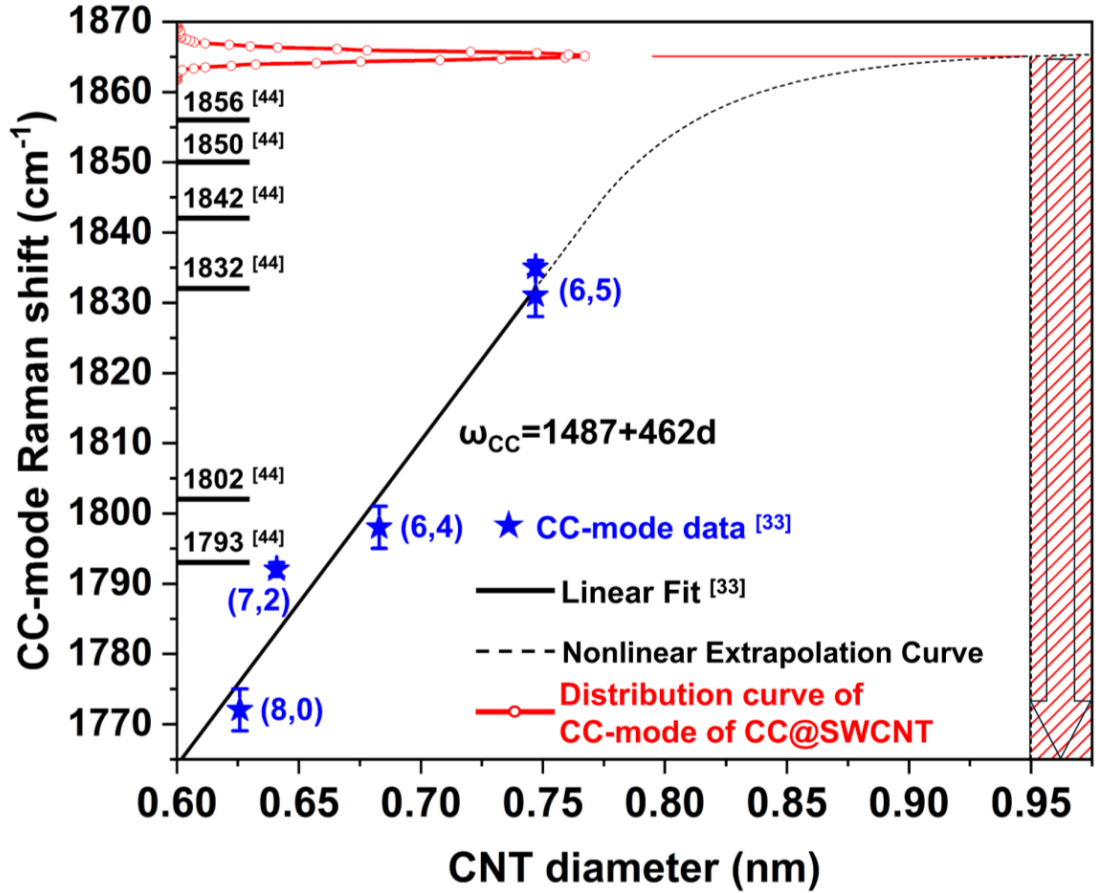

**Figure 6.** The CC-mode Raman shift ($\omega_{cc}$) as a function of CNT host diameter. Experimental CC-mode data (blue stars) is fitted linearly (solid black line) using the equation $\omega_{cc}=1487+462d$, where d is the CNT diameter cited from previous paper [33]. A nonlinear extrapolation curve (dashed black line) extends the trend beyond the linear regime. The red curve represents the distribution of CC-mode Raman shifts for CC@SWCNT, highlighting the deviation from linearity in the large-diameter range. The shaded red area denotes the range of large-diameter SWCNTs suitable for CC@SWCNT synthesis.

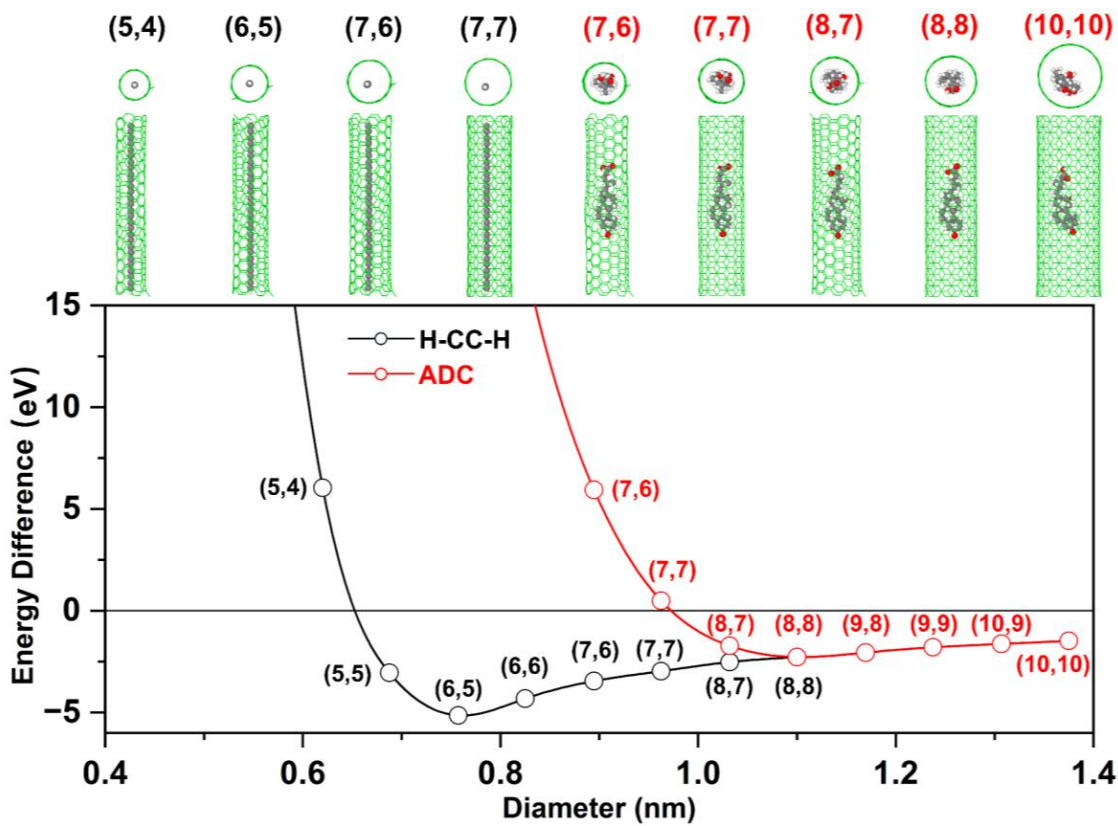

**Figure 7.** Energy difference as a function of SWCNT diameter for two configurations: H-CC-H (black line) and ADC (red line). The ADC and CC encapsulated within SWCNTs of specific diameters are visualized above the plot. Cross-sectional views are included to illustrate the interaction between the ADC/CC configurations and the SWCNT walls.

Lastly, we studied the thermal stability of CC@SWCNT, which is a critical parameter essential for the subsequent synthesis of materials and fabrication of devices based on CC@SWCNT. According to previous research, $C_{10}H_2$@SWCNT lost its Raman signal due to polyyne when the temperature increased to 500°C, indicating that the trapped $C_{10}H_2$ molecules were liberated from the SWCNTs under heating treatment [55]. In 2020, we successfully synthesized BN and $MoS_2$ layers on the exterior of SWCNTs, constructing a SWCNT@BN@$MoS_2$ 1D heterostructure [56], allowing us to explore the possibility of synthesizing a CC@SWCNT@BN structure, further enriching the family of 1D heterojunction materials and investigating the potential impact of CC on the overall structure. However, this requires thermal stability of CC inside the SWCNT, as the BN synthesis process requires temperatures exceeding 1000°C. To clarify this, we used CC@SWCNT as the substrate for BN synthesis at 1075°C. Surprisingly, Raman spectroscopy showed that the CC-mode signal still exists after BN synthesis, albeit with reduced intensity (Fig. S12). In the process of synthesizing SWCNT@BN using SWCNT as the substrate, we also observed a reduction in the intensity of the G band. This reduction is likely because of the loss of some SWCNTs due to the high temperatures during the synthesis process. A similar decrease in CC-mode intensity should be expected. However, no detachment of CC was found, unlike for the polyyne@SWCNT at 500°C. This indicates that CC@SWCNT exhibits more excellent thermal stability than the SWCNT materials filled with organic molecules.

**CONCLUSION**

In this study, we successfully synthesized confined carbyne (CC) within single-walled carbon nanotubes (SWCNTs) using surfactant molecules (ADC) as precursors. This approach marks the first utilization of surfactants in the synthesis of CC@SWCNT, simplifying the process and enhancing the synthesis efficiency of CC. Notably, the synthesis was achieved at a remarkably low temperature of 400°C, the lowest reported temperature for CC@CNT synthesis to date, which minimizes potential damage to the SWCNT structure. Furthermore, experiment and molecular simulation results indicate that the SWCNTs encapsulating CC must possess large diameters, exceeding 0.95 nm, a notable increase from previously reported optimal diameters for CC synthesis (~0.7 nm), thus resulting in a higher CC-mode Raman shift (1863–1867 cm$^{-1}$). The larger tube diameter reduces the interaction between CC and SWCNT, therefore forming "weakly confined carbyne". Moreover, we found that the linear relationship between the CC mode and the CNT host is only applicable within a certain range, and it deviates from linearity (saturates) in our samples, consistent with the saturation of host-CC interaction in the quasi-free standing, yet confined, carbyne limit. This method successfully filled the gap in synthesizing CC inside large-diameter SWCNT (over 0.95 nm).

**METHODS**

**SWCNT dispersion and surfactant preparation**

HiPco SWCNT (NanoIntegris, Batch# R2-172) was dispersed by tip-sonication method, followed by liquid chromatography to separate semi-conducting and metallic SWCNT dispersion[34]. The dispersion was finally replaced with 0.5% sodium cholate (SC) by the cross-flow method. Sodium deoxycholate (DOC) (Fujifilm Wako Pure Chemical, ≥ 96.0%), sodium dodecyl sulfate (SDS) (Fujifilm Wako Pure Chemical, ≥ 95.0%), and sodium cholate (SC) (Fujifilm Wako Pure Chemical, ≥ 98.5%) were purchased from a commercial source. Solutions of 1 wt% DOC, SDS, and SC were prepared by dissolving 2g DOC, SDS, and SC with 198 g deionized water. 1 wt% ammonium deoxycholate (ADC) solution (200 mL) was prepared by dissolving 2 g deoxycholate acid (Sigma-Aldrich, ≥ 99.0%) with 98 g deionized water, followed by adding 1 mL of ammonium hydroxide (28% $NH_3$ in $H_2O$, Sigma-Aldrich), then diluted by 99 g of deionized water.

**Surfactant exchange in SWCNT dispersion**

Ultrafiltration facilitated the surfactant exchange process from SC to DOC, SDS, or ADC. Initially, 3 mL of SWCNT-SC dispersion was mixed with 9 mL of a 1 wt% DOC, SDS, or ADC solution in an Amicon® Ultra-15 Centrifugal Filter Unit (100 kDa). The mixture was then centrifuged at 2,000 rpm for 5 minutes. During this process, the SWCNTs in solution were retained in the inner tube, while the surfactant solution permeated through the membrane into the outer tube and was subsequently discarded. 10 mL of fresh 1 wt% DOC, SDS, or ADC solution was added to the inner tube, followed by centrifugation at 2,000 rpm for 5 minutes. The replacement cycles were carried out for 10 iterations, resulting in an estimated dilution of the first surfactant by a factor of $1.65 \times 10^{-8}$. After completing the final cycle, the remaining approximately 2 mL of SWCNT dispersion in the inner tube was diluted to 6 mL with 1 wt% DOC or SDS or ADC solution.

**SWCNT film fabrication**

After the surfactant exchange process, the SWCNT dispersion was used to fabricate SWCNT films. An AAO membrane [35] (Whatman, 25 mm diameter, 0.02 μm pore size) was positioned on a fritted glass base attached to a filter flask connected to a vacuum pump. Approximately 1 mL of deionized water was dropped onto the AAO membrane to thoroughly wet its surface, followed by vacuum filtration to remove the excess water. Subsequently, 500 μL of the SWCNT solution, which had the highest optical absorbance of 0.286 at 1192 nm in the $E_{11}$ region (Fig. S13), was added to the moistened AAO membrane drop by drop using a pipette. A thin SWCNT film formed on the AAO membrane.

The AAO membrane was then floated on an HCl bath and heated on a hot plate at 60°C for 20 minutes to detach the SWCNT film from the membrane. The SWCNT film was released by immersing the membrane in deionized water and retrieved using a perforated silicon substrate (hole size 100μm × 100μm). The SWCNTs on the silicon substrate were then heated on a hot plate at 80°C to gradually remove excess water from the film.

**CC synthesis process**

The SWCNT on the silicon substrate was annealed in a CVD chamber at 400°C under an argon gas atmosphere. The detailed parameters for the annealing process are as follows: the temperature was raised to 400°C in 1 hour and 15 minutes, followed by maintaining at 400°C for 1 hour and 30 minutes. The argon gas flow was maintained at 300 sccm, and the pressure was kept within the 330–350 Pa range. After heating, the system was allowed to cool naturally, completing the synthesis of CC@SWCNT.

**BN synthesis process**

The CC@SWCNT@BN was prepared using the CVD method. The CC@SWCNT sample was placed at the center of the furnace. 30 mg of ammonia−borane complex (97%, Sigma-Aldrich) was loaded upstream and heated to 70 °C. The vapor of the BN precursor was carried by a flow of 200 sccm Ar (with 3% $H_2$) to the heating zone. The reaction temperature was 1075 °C, and the growth time was 60 minutes.

**Raman characterization**

The Raman spectra of CC@SWCNT were measured using Renishaw Raman spectroscopy equipment. Various laser wavelengths and diffraction gratings—488 nm (2400 grooves/mm), 532 nm (1800 grooves/mm), 633 nm (1800 grooves/mm), and 785 nm (1200 grooves/mm)—were employed to assess the intensity and wavenumber of the characteristic CC peaks. All measurements were performed in a backscattering geometry utilizing a 50× long-distance objective lens under ambient conditions, and the laser power used during the measurement was lower than 1 mW to prevent potential heating effects on CC@SWCNT samples. Raman mapping was conducted across the CC@SWCNT film. The scanning area for Raman mapping was a 45 μm × 45 μm square, with a step size of 3 μm. A total of 961 points (31 x 31) were scanned for each sample.

**Temperature-dependent Raman characterization**

The temperature-dependent Raman characterization was measured using the THMS600 heating and freezing stage. The CC@SWCNT sample was loaded into the sample chamber, and liquid nitrogen was introduced into the chamber for cooling. The sample was first cooled to -195°C, and then slowly heated at 50°C/min, with characterization conducted at -165°C, -140°C, -115°C, -90°C, -65°C, -40°C, -15°C, 10°C, 35°C, 60°C, and 85°C. The laser wavelength and diffraction grating were 532 nm and 2400 grooves/mm, with power levels of 1% (0.324 mW) and 3.2% (1.357 mW).

**Supporting Information**.

The Supporting Information is available free of charge at

S1. CC-mode and Corresponding Resonant Excitation Laser Energy of CC@DWCNT and CC@SWCNT

S2. Statistical Analysis of RBM and G Peak Fitting Results for Pristine SWCNT, Annealed SWCNT, and CC@SWCNT

S3. Scanning Electron Microscope (SEM) of Semi-conducting HiPco SWCNT Film Before and After Annealing

S4. Raman Mapping and Normalized CC-Mode Intensity ($I_{CC}/I_{G+}$) Distribution Curve for CC@SWCNT Samples Synthesized by 400°C Annealing Treatment for 1.5 Hours (Sample A-G)

S5. Average Raman Spectra by Averaging All the Spectra Obtained at Different Points in the Raman Mapping of CC@SWCNT (Sample A-G)

S6. Raman Mapping, Normalized CC-Mode Intensity ($I_{CC}/I_{G+}$) Distribution Curve, and Average Raman Spectra of CC@SWCNT Synthesized from SC, DOC, SDS, and ADC

S7. Raman Mapping and Average Raman Spectrum of CC@SWCNT Synthesized from ACCVD SWCNTs

S8. Photoluminescence Excitation (PLE) Mapping and Chirality Distribution Mapping of ACCVD and Semi-Conducting HiPco SWCNTs

S9. Relationship between SWCNT Diameter and Deviation of Encapsulated Carbyne from the Center

S10. Average Raman spectra of SWCNTs annealed under different temperatures (300°C, 350°C, 400°C, 500°C)

S11. Temperature-dependent Raman characterization of CC@SWCNT under different laser energy

S12. Raman Spectra of CC@SWCNT before and after BN growth

S13. Absorbance Spectra of SWCNT Dispersion Before and After Ultrafiltration Process

**AUTHOR INFORMATION**

Corresponding Author

**Shigeo Maruyama** − Department of Mechanical Engineering, The University of Tokyo, Tokyo 113-8656, Japan; orcid.org/0000-0003-3694-3070; Email: maruyama@photon.t.u-tokyo.ac.jp

Authors

**Bo-wen Zhang**−Department of Mechanical Engineering, The University of Tokyo, Tokyo 113-8656, Japan

**Xi-yang Qiu**−Department of Mechanical Engineering, The University of Tokyo, Tokyo 113-8656, Japan

**Yicheng Ma**−State Key Laboratory of Fluid Power and Mechatronic Systems, School of Mechanical Engineering, Zhejiang University, Hangzhou 310027, People's Republic of China

**Qingmei Hu**−Department of Mechanical Engineering, The University of Tokyo, Tokyo 113-8656, Japan

**Aina Fitó-Parera**−Theory and Spectroscopy of Molecules and Materials, Department of Physics, University of Antwerp, Antwerp 2610, Belgium

**Ikuma Kohata**−Department of Mechanical Engineering, The University of Tokyo, Tokyo 113-8656, Japan

**Ya Feng**−Key Laboratory of Ocean Energy Utilization and Energy Conservation of Ministry of Education, School of Energy and Power Engineering, Dalian University of Technology, Liaoning 116024, China

**Yongjia Zheng**−Department of Mechanical Engineering, The University of Tokyo, Tokyo 113-8656, Japan; State Key Laboratory of Fluid Power and Mechatronic Systems, School of Mechanical Engineering, Zhejiang University, Hangzhou 310027, People's Republic of China; orcid.org/0000-0001-5836-6978

**Chiyu Zhang**−Department of Chemistry and Biochemistry, Chemical Physics Program, and Maryland Nano Center, University of Maryland, College Park, Maryland 20742, United States; orcid.org/0000-0002-5664-1849

**Yutaka Matsuo**−Department of Chemical Systems Engineering, Graduate School of Engineering, Nagoya University, Nagoya 464-8603, Japan; Institute of Materials Innovation, Institutes for Future Society, Nagoya University, Nagoya 464-8603, Japan; orcid.org/0000-0001-9084-9670

**YuHuang Wang**−Department of Chemistry and Biochemistry, Chemical Physics Program, and Maryland Nano Center, University of Maryland, College Park, Maryland 20742, United States; orcid.org/0000-0002-5664-1849

**Shohei Chiashi**−Department of Mechanical Engineering, The University of Tokyo, Tokyo 113-8656, Japan; orcid.org/0000-0002-3813-0041

**Keigo Otsuka**−Department of Mechanical Engineering, The University of Tokyo, Tokyo 113-8656, Japan; orcid.org/0000-0002-6694-0738

**Rong Xiang**− Department of Mechanical Engineering, The University of Tokyo, Tokyo 113-8656, Japan; State Key Laboratory of Fluid Power and Mechatronic Systems, School of Mechanical Engineering, Zhejiang University, Hangzhou 310027, People's Republic of China; orcid.org/0000-0002-4775-4948

**Dmitry I. Levshov**−Department of Mechanical Engineering, The University of Tokyo, Tokyo 113-8656, Japan; Theory and Spectroscopy of Molecules and Materials, Department of Chemistry, University of Antwerp, Antwerp 2610, Belgium; orcid.org/0000-0002-2249-7172

**Sofie Cambré**−Theory and Spectroscopy of Molecules and Materials, Department of Physics, University of Antwerp, Antwerp 2610, Belgium; orcid.org/0000-0001-7471-7678

**Wim Wenseleers**−Nanostructured and Organic Optical and Electronic Materials, Department of Physics, University of Antwerp, Antwerp 2610, Belgium

**Slava V. Rotkin**−Materials Research Institute Department of Engineering Science & Mechanics, The Pennsylvania State University, University Park, Pennsylvania16802, United States, orcid.org/0000-0001-7221-1091

Acknowledgment

The authors thank Professor Hiromichi Kataura and Takeshi Tanaka for providing the HiPco SWCNT dispersion.

Funding Sources

This research was financially supported by the Japan Society for the Promotion of Science (JSPS) KAKENHI under Grant Numbers JP23H00174, JP23H05443, and JP21KK0087, as well as by JST CREST under Grant Number JPMJCR20B5, Japan.

Additionally, this work was supported by a JSPS-FWO Bilateral Joint Research Project (Grant Numbers: JSPS JPJSBP120212301 and FWO VS08521N).

Financial support was also provided by JST SPRING under Grant Number JPMJSP2108.

Notes

The authors declare no competing financial interest.

**ABBREVIATIONS**

CC: Confined Carbyne

wCC: Weakly Confined Carbyne

CNT: Carbon Nanotube

MWCNT: Multi-Walled Carbon Nanotube

DWCNT: Double-Walled Carbon Nanotube

SWCNT: Single-Walled Carbon Nanotube

HTHV: High-Temperature High-Vacuum

BLA: Bond-Length Alternation

ADC: Ammonium Deoxycholate

DOC: Sodium Deoxycholate

SC: Sodium Cholate

SDS: Sodium Dodecyl Sulfate

CVD: Chemical Vapor Deposition

ACCVD: Alcohol Catalytic Chemical Vapor Deposition